\documentclass[11pt]{article}

\usepackage[margin=1in]{geometry}
\usepackage[utf8]{inputenc}
\usepackage[T1]{fontenc}
\usepackage{amsmath,amssymb,amsfonts,amsthm}
\usepackage{bm}
\usepackage{bbm}
\usepackage{graphicx}
\usepackage{subcaption}
\usepackage{float}
\usepackage{xcolor}
\usepackage{booktabs}
\usepackage{algorithm}
\usepackage[noend]{algpseudocode}
\usepackage{tikz}
\usetikzlibrary{positioning,arrows.meta,shapes.geometric}
\usepackage[authoryear,round]{natbib}
\usepackage{authblk}
\usepackage[colorlinks=true,citecolor=blue,urlcolor=blue,linkcolor=blue]{hyperref}

\DeclareUnicodeCharacter{00B0}{\ensuremath{^\circ}}
\DeclareUnicodeCharacter{2013}{--}
\DeclareUnicodeCharacter{2014}{---}
\DeclareUnicodeCharacter{2019}{'}
\DeclareUnicodeCharacter{201C}{``}
\DeclareUnicodeCharacter{201D}{''}

\theoremstyle{plain}

\newtheorem{theorem}{Theorem}[section]

\newtheorem{Proposition}{Proposition}[section]

\theoremstyle{definition}
\newtheorem{definition}[theorem]{Definition}
\newtheorem{assumption}{Assumption}[section]

\newcommand{\bu}{\mathbf{ u}}
\newcommand{\bv}{\mathbf{ v}}

\newcommand{\bx}{\mathbf{ x}}

\newcommand{\bz}{\mathbf{ z}}
\newcommand{\bA}{\mathbf{ A}}

\newcommand{\bE}{\mathbf{ E}}

\newcommand{\bK}{\mathbf{ K}}

\newcommand{\bX}{\mathbf{ X}}

\newcommand{\bY}{\mathbf{ Y}}
\newcommand{\bZ}{\mathbf{ Z}}

\newcommand{\bbeta}{\bm{\beta}}

\newcommand{\bgamma}{\bm{\gamma}}

\newcommand{\bnu}{\bm{\nu}}

\newcommand{\bSigma}{\bm{\Sigma}}

\newcommand{\bOmega}{\boldsymbol{\Omega}}
\begin{document}

\title{Stein-Encoder: A White-Box Supervised Encoder via Stein Identities in Multi-Modal Studies}



\author[1]{Jiarui Zhang\textsuperscript{*}}
\author[2]{Shuoxun Xu\textsuperscript{*}}
\author[3]{Jiasheng Shi\textsuperscript{$\dagger$}}
\author[4]{Xinzhou Guo\textsuperscript{$\dagger$}}

\affil[1]{School of Mathematics, South China University of Technology, Guangzhou, China}
\affil[2]{Department of Biostatistics and Epidemiology, University of California, Berkeley, CA, USA}
\affil[3]{School of Data Science, The Chinese University of Hong Kong, Shenzhen, China}
\affil[4]{Department of Mathematics, Hong Kong University of Science and Technology, Hong Kong, China}

\date{}

\maketitle

\footnotetext[1]{\textsuperscript{*} Jiarui Zhang and Shuoxun Xu contributed equally to this work.}
\footnotetext[2]{\textsuperscript{$\dagger$} Co-corresponding authors: Jiasheng Shi (shijiasheng@cuhk.edu.cn), Xinzhou Guo (xinzhoug@ust.hk).}

\maketitle

\begin{abstract}
In multi-modal biomedical research, integrating high-dimensional genomic data with clinical baselines is essential for precision medicine. However, standard deep neural network approaches often entangle these modalities, obscuring the specific predictive impact of genetic features and leading to possibly suboptimal predictive performance. Motivated by the landmark METABRIC cohort primary breast tumors study, we propose the Stein-Encoder, a white-box supervised framework designed to isolate the genetic signal driving clinical outcomes conditional on nuisance covariates. By leveraging Stein's method and residualization techniques, our approach constructs an interpretable single index that summarizes relevant biological heterogeneity while flexibly incorporating clinical factors and can be used to improve downstream prediction. We establish theoretical guarantees for identification, consistency and efficiency improvement. Applied to the METABRIC cohort, the Stein-Encoder outperforms unsupervised benchmarks in predictive accuracy. Crucially, it achieves structural disentanglement by revealing response-specific biological mechanisms: we find that tumor size is driven primarily by mitotic networks, whereas prognostic indices rely on a distinct proliferation-versus-immune axis. This work contributes a unified, computationally efficient framework that bridges statistical rigor with the representational power of neural networks, enabling interpretable, task-specific and efficient compression of multi-modal health data for a wide range of precision medicine applications, beyond biomarker discovery.
\end{abstract}

\bigskip

\section{Introduction}
\label{sec:intro}
\subsection{Motivations and Objectives}\label{sec:motivation}

The rapid development of Artificial Intelligence (AI), particularly the emergence of Deep Neural Network (DNN), has fundamentally transformed data-driven discovery across various scientific disciplines, exemplified by the accurate prediction of 3D protein structures via AlphaFold \citep{jumper2021highly} and the accelerated discovery of crystals in materials science \citep{merchant2023scaling}. While the application of DNN has achieved remarkable success in scientific discoveries, how to achieve dual requirements of prediction accuracy and interpretability with DNN, a black-box method, remains a critical challenge. This question is particularly important in multi-modal studies where researchers are not only interested in utilizing different modalities to improve outcome prediction accuracy but also in interpreting the impact of a certain modality of interest \citep{lecun2015deep}. 


A motivating example of this paper is the METABRIC study \citep{curtis2012genomic}, a multi-modal cohort of breast cancer. In the METABRIC study, two scientific tasks are of great interest, (1) predict key clinical outcomes based on two different modalities, clinical biomarkers and high-dimensional gene expression profiles and (2) interpret the impact of gene expression modality on clinical outcomes \citep{kristensen2014principles}. The first task integrates clinical baselines with genomic profiles to capture patient heterogeneity beyond uni-modal assessments, achieving more accurate clinical prediction \citep{hasin2017multi}.  The second task aims to isolate actionable genetic drivers, ensuring the clinical transparency required for high-stakes medical decision-making \citep{rudin2019stop}. Simply using  standard end-to-end DNNs to fuse the two modalities may deliver neither optimal predictive performance nor desired interpretability \citep{ching2018opportunities}. First, in the high-dimensional regime, these over-parameterized models are prone to overfitting stochastic genomic noise rather than learning robust predictive signals, and may suffer suboptimal prediction \citep{wainberg2018deep}. Second, DNN tends to entangle the strong, baseline signals from clinical variables with the subtle, high-dimensional signals from gene expression in complex nonlinear ways, and cannot inform practitioners how gene expression profiles impact clinical outcomes.

To address the two scientific tasks in the METABRIC study, or more broadly, multi-modal studies, the key is to construct a white-box encoder which is interpretable and sufficiently compresses the modality of interest. In AI, an encoder refers to a mapping from raw inputs into a low-dimensional representation and is typically embedded into DNN \citep{goodfellow2016deep}. In the community of AI, numerous encoders have been proposed. For example, Variational Autoencoders (VAE) \citep{kingma2013auto} and Denoising Autoencoders (DAE) \citep{vincent2008extracting} are widely utilized to learn robust latent representations. However, AI-based encoders are typically black-box and focus on improvement of general prediction ability. The compressed values of predictors from AI-based encoders fail to disentangle the incremental predictive information of the genetic modality from the baseline effects of the nuisance clinical covariates and are thus not interpretable in the METABRIC study. Moreover, the prediction ability of AI-based encoders may not be optimal in the METABRIC study due to their unsupervised nature. In sum, the AI-based encoder is not suitable for the scientific tasks considered in the METABRIC study and may not help achieve the dual requirements of prediction accuracy and interpretability in multi-modal studies. 

In this paper, we aim to utilize appropriate statistical principle to define a white-box encoder in multi-modal studies, specifically for the METABRIC study with high-dimensional genetic features. To define encoders, several statistical principles have been considered. 
Based on Principal Component Analysis,  several encoders have been proposed but they are unsupervised and prioritize variance over predictive relevance \citep{wainberg2018deep}. Based on Sufficient Dimension Reduction (SDR), Sliced Inverse Regression (SIR) \citep{li1991sliced} and Sliced Average Variance Estimation (SAVE) \citep{cook1991sliced} target the predictive subspace but struggle to handle the conditioning on complex, mixed-type nuisance modalities effectively. Therefore, a new statistical principle which can account for different modalities in the supervised setting is needed. Ultimately, we wish the newly defined white-box encoder can enhance prediction accuracy and interpretability of DNN in the METABRIC study.

\subsection{An Overview of Research Methods and Findings}\label{sec:overview}

\begin{figure}[H]
    \centering
    \resizebox{0.75\linewidth}{!}{%
    \begin{tikzpicture}[
        node distance = 6mm and 20mm,
        every node/.style = {font=\footnotesize},
        data/.style = {
            cylinder,
            shape border rotate=90,
            aspect=0.28,
            draw,
            minimum height=12mm,
            minimum width=14mm,
            align=center,
            cylinder body fill=white,
            cylinder end fill=white,
            inner sep=1pt
        },
        process/.style = {
            rectangle,
            rounded corners=2pt,
            draw,
            fill=blue!5,
            minimum width=34mm,
            minimum height=8mm,
            align=center,
            inner sep=2pt
        },
        encoder/.style = {
            trapezium,
            trapezium left angle=70,
            trapezium right angle=110,
            draw,
            fill=teal!8,
            minimum width=34mm,
            minimum height=8mm,
            align=center,
            inner sep=2pt
        },
        dnn/.style = {
            rectangle,
            rounded corners=3pt,
            draw,
            very thick,
            fill=orange!10,
            minimum width=36mm,
            minimum height=10mm,
            align=center,
            inner sep=2pt
        },
        arrow/.style = {->, >=Stealth, thick},
        baselineArrow/.style = {->, >=Stealth, thick, dashed}
    ]


    \node[data] (data) {Raw\\data\\$(Y,\mathbf{X},\mathbf{Z})$};

    \node[process, below=6mm of data] (step1) {%
        \textbf{Step 1: Nuisance removal}\\
        Fit $\mathbf Z\mid \mathbf X \sim \mathcal{N}(\mathbf A\mathbf X,\mathbf \Sigma)$,\\
        residuals $\mathbf{Z}' = \mathbf{Z} - \widehat{\mathbf A}\mathbf X$
    };

    \node[encoder, below=6mm of step1] (encoder) {%
        \textbf{Step 2: Stein-Encoder}\\
        Stein identities with probes $T(Y)$,\\
        estimate $\widehat{\bgamma}$, score $t=\widehat{\bgamma}^\top \mathbf Z$
    };

    \node[dnn, right=30mm of step1] (dnn) {%
        \textbf{Downstream DNN $f_{\Theta}(\cdot)$}\\
        same architecture,\\
        different inputs for each path
    };


    \draw[arrow]
        (data) -- node[left,pos=0.4]{\scriptsize proposed path}
        (step1);

    \draw[arrow]
        (step1) -- (encoder);

    \draw[arrow]
        (encoder.east) -|
        node[above,pos=0.35]{\scriptsize input $(\mathbf X,\,\widehat{\bgamma}^\top \mathbf Z)$}
        (dnn.south);


    \draw[baselineArrow]
        (data.east) -|
        node[above,pos=0.35]{\scriptsize traditional path, input $(\mathbf X,\mathbf Z)$}
        (dnn.north);

    \end{tikzpicture}%
    }
    \caption{Data analysis pipeline.}
    \label{fig:pipeline_stein_encoder}
\end{figure}
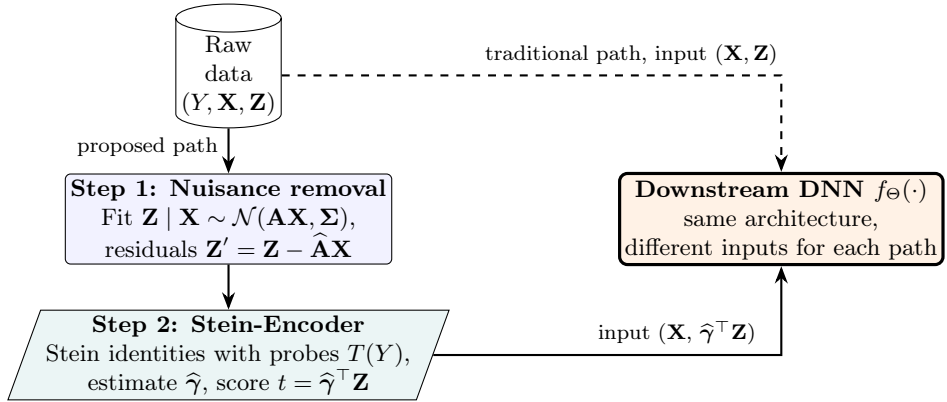

To achieve accurate prediction and transparent interpretation in the METABRIC study, we propose a new data analysis pipeline as illustrated in Figure \ref{fig:pipeline_stein_encoder}. The new data analysis pipeline is based on a novel white-box encoder constructed by Stein's identity, a statistical principle widely used in distributional approximation and variational inference \citep{stein1981estimation}. Specifically, we propose a Stein-Encoder to summarize and interpret the impact of genetic expressions modality to cancer-related outcomes, and apply DNN to the encoder and clinical modality to predict the outcomes.  Unlike classical encoders, the Stein-Encoder is a white-box encoder capturing the intrinsic structural information of the modality of interest, genetic expression profile, while rigorously accounting for the context provided by the auxiliary modality, clinical biomarkers. Under mild assumptions, we show that the Stein-Encoder provides a single index which can be interpreted as a conditional genetic risk score and serves as a statistically optimal "pre-training" step which effectively reduces the dimensionality of the problem without information loss for downstream DNN \citep{rudin2019stop}. The estimation consistency of the Stein-Encoder is also established in both low and high dimensions.

With the proposed Stein-Encoder, the analysis of the multi-modal METABRIC breast cancer study is substantially facilitated. Specifically, we find that the Stein-Encoder significantly outperforms the unsupervised dimensionality reduction method PCA and standard neural networks in predicting key cancer-related outcomes. In the METABRIC study, the Stein-Encoder method reduces test MSE for tumor size from 140.5 (standard DNN) to 126.2 and increases $R$-squared for age at diagnosis from 0.59 to 0.64. In addition, compared with PCA, the learned genetic index from Stein-Encoder exhibits a clearer monotone gradient in tumor size, NPI and nodal status and is enriched for proliferation and immune-related genes. This reveals a biologically coherent transcriptomic risk axis that PCA and black-box models fail to isolate. Moreover, we distinguish response-specific transcriptomic drivers, isolating a pure mitotic network (e.g., \textit{AURKB}, \textit{FOXM1}) for tumor size while recovering a ``proliferation vs. immune'' axis (e.g., \textit{CD79A}) for NPI.  This biological disentanglement contrasts with standard PCA, which captures only generic proliferation signals regardless of the clinical endpoint, demonstrating our method's ability to uncover outcome-relevant heterogeneity.

In summary, this work contributes a unified framework for multi-modal representation learning that bridges the gap between statistical rigor and representational power. 
We provide a closed-form, computationally efficient estimator that acts as a plug-and-play module to enhance the performance of downstream neural networks by reducing input dimensionality and isolating relevant heterogeneity. 
By offering a transparent mechanism to fuse and interpret complex multi-modal data transforming raw associations into actionable biological insights, the Stein-Encoder advances the methodology for precision medicine and fosters the clinical trust required for deploying AI in high-stakes scientific domains, particularly the METABRIC study.

\textbf{Organization of the Paper.} The remainder of this paper is organized as follows. Section \ref{sec:data_problem} introduces the data and scientific task based on the METABRIC cohort and formalizes the multi-modal problem setup. Section \ref{sec:methodology} details the Stein-Encoder methodology, defining and interpreting the encoder, tailoring the encoder to the METABRIC study and deriving the unified estimation algorithm and use of the Stein-Encoder. Section \ref{sec:theory} presents the theoretical properties, including identification, estimation consistency and asymptotic efficiency gain  of Stein-Encoder. Section \ref{sec:simulation} evaluates the method's performance through extensive numerical experiments. Section \ref{sec:realdata} demonstrates the practical application of the Stein-Encoder to real-world cancer data, comparing it against benchmark methods. Finally, Section \ref{sec:discussion} concludes the paper with a discussion of implications and future research directions.

\textbf{Notation.} Throughout the paper, we use $\mathbb{E}[\cdot]$ to denote the expectation operator and $\nabla$ for the gradient operator. For a vector $\mathbf{v}$, $\|\mathbf{v}\|_2$ denotes the Euclidean norm, and for high-dimensional analysis, $\|\mathbf{v}\|_0$ denotes the sparsity level (number of non-zero elements). We denote $\gamma^{\otimes k} = \underbrace{\gamma \otimes \gamma \otimes \cdots \otimes \gamma}_{k \text{ times}}$ as the $k$-th Kronecker power of the vector $\gamma$.

\section{Data, Scientific Tasks and Problem Setup}\label{sec:data_problem}
In this section, we first introduce the METABRIC study and the scientific tasks in this multi-modal study in Section 2.1. Then, we mathematically define the problem setup, illustrate the limitations of existing methods and discuss what type of encoder we seek in the multi-modal study in Section 2.2.

\subsection{Data and Scientific Tasks}\label{sec:study_design}
In this paper, we consider the Molecular Taxonomy of Breast Cancer International Consortium (METABRIC) cohort \citep{curtis2012genomic} as the motivating example, which represents one of the largest and most comprehensively annotated collections of primary breast cancer specimens. In our analysis, after standard quality control procedures-specifically excluding subjects with missing response values and selecting the top 400 gene probes ranked by marginal variance, we retain approximately 1900 patients with complete data, and consider the following two modalities, (1) a nuisance modality with about 400 dimensions collecting several dozen clinical and copy-number covariates and (2) a high-dimensional genetic modality consisting of 400 highly variable gene expression probes from the Illumina HT-12 v3 platform. Unlike standard unimodal datasets, this cohort presents a complex multi-modal structure involving clinical phenotypes, structural genomic variations, and transcriptomic profiles. {\color{black}The scientific task of the study is to (1) predict cancer-related outcomes based on the two modalities and (2) interpret the impact of the genetic modality. The details of the modalities and cancer-related outcomes, and practical importance of the tasks are provided as follow.}

The first set of the study is denoted as the {nuisance modality} ($\bX$). In the context of semiparametric statistical inference, a "nuisance" parameter refers to a variable that is not the primary target of inference but must be accounted for to obtain unbiased estimates of the target parameter \citep{robinson1988root,chernozhukov2018double}. Here, $\bX$ comprises low-dimensional variables defining the heterogeneous biological background, including standard clinico-pathological factors (e.g., treatment status, tumor grade, and estrogen receptor status). Crucially, we also include Copy Number Aberrations (CNA) in this nuisance category. Although genomic in origin, CNAs represent discrete structural events (encoded as integers $\{-2, -1, 0, 1, 2\}$) that drive baseline cellular states. As noted in integrative genomic analyses \citep{kristensen2014principles}, controlling for these structural variations is essential to distinguish the functional regulation of mRNA expression from simple gene dosage effects. The second set of the study constitutes the {primary feature modality} ($\bZ$), consisting of high-dimensional continuous gene expression profiles. We focus on a subset of genes exhibiting significant variability, serving as the candidate features for identifying functional drivers of disease progression. We analyze four clinical outcomes ($Y$): the Nottingham Prognostic Index (NPI) \citep{galea1992nottingham} and its individual components—tumor size and lymph node status—to disentangle their specific genetic associations. Additionally, we include age at diagnosis to capture age-dependent heterogeneity in tumor biology \citep{partridge2016subtype}.

{\color{black}The practical importance of the study is two-fold.} First, we will predict the clinical outcomes ($Y$) by integrating the nuisance clinical background ($\bX$) and the genetic information ($\bZ$). Such prediction is important in the era of precision medicine because it moves beyond coarse anatomical staging to capture the underlying molecular heterogeneity, enabling more granular risk stratification and the tailoring of therapeutic interventions to individual biological profiles. Second, we will determine whether the high-dimensional genome ($\bZ$) adds independent predictive value over the established clinical and structural factors ($\bX$). While clinical variables provide a baseline for risk stratification, the core scientific question is to identify an interpretable index of gene expression that drives clinical outcomes conditional on the heterogeneous background. Isolating this signal is essential for discovering valid biomarkers that are not merely surrogates for known clinical confounders.


\subsection{Problem Setup}\label{sec:problem_setup}
We formalize the multi-modal prediction problem using the following mathematical notation. 
Let $Y \in \mathbb{R}$ denote the scalar response variable. 
The feature space consists of two distinct components: a vector of nuisance covariates $\bX \in \mathbb{R}^p$, and a vector of genetic features $\bZ \in \mathbb{R}^q$. 
Mathematically, the scientific objectives of this study translate into two distinct statistical challenges: (1) to learn the joint conditional expectation $\mathbb E[Y|\bX,\bZ]$ that minimizes generalization error; and (2) to identify the specific structural component of $\bZ$ that drives variation in $Y$ conditional on $\bX$, thereby disentangling the genetic impact from the confounding clinical baseline.
In its most general form, without imposing specific structural assumptions, the data generating process encompassing these objectives can be expressed as:
\begin{equation}
    \label{eq:general_model}
    Y = g(\bX, \bZ) + \varepsilon,
\end{equation}
where $\varepsilon$ is a noise term independent of $\bX$ and $\bZ$ and $g(\cdot)$ is an unknown, potentially complex function capturing the underlying biological mechanism.

Modern artificial intelligence (AI) approaches, particularly multi-modal DNN, offer powerful tools for approximating such complex functions. A direct approach typically involves concatenating $\bX$ and $\bZ$ into a single input vector for a DNN. While these "black-box" models often achieve high predictive accuracy by flexibly approximating $g$, they suffer from significant limitations in scientific discovery. The primary issue is the {entanglement} of features: such models tend to mix the strong, baseline signals from clinical variables $\bX$ with the subtle, high-dimensional signals from $\bZ$ in complex nonlinear ways \citep{baltruvsaitis2018multimodal}. Consequently, it becomes nearly impossible to disentangle the specific contribution of the genetic modality from the clinical background. Moreover, the prediction might not be optimal as these over-parameterized models are prone to overfitting stochastic genomic noise rather than learning robust predictive signals \citep{wainberg2018deep}.  As emphasized by \citep{rudin2019stop}, for high-stakes decisions in healthcare, reliance on opaque models is problematic; distinct and interpretable attribution of risk factors is required.

To address this interpretability challenge while maintaining predictive power, we propose to structure the genetic component through a dimension reduction perspective. Specifically, our goal is to identify a predictive and interpretable {linear encoder} for the genetic features, denoted as $\bgamma^\top \bZ$, where $\bgamma \in \mathbb{R}^q$ is an unknown projection direction. By imposing this structure, we aim to conduct prediction and approximate the general relationship in \eqref{eq:general_model} with DNN regressing $Y$ on $\bX$ and $\bgamma^\top \bZ$ rather than $\bX$ and $\bZ$ directly.
This choice of a linear encoder is theoretically grounded in quantitative genetics, where complex traits are often effectively modeled as the additive effects of many variants—a principle underpinning the success of Polygenic Risk Scores (PRS) \citep{visscher201710}. The weights in $\bgamma$ thus provide a direct, interpretable measure of gene importance, striking a balance between biological plausibility and the statistical flexibility required to model clinical covariates.

However, determining an interpretable and predictive linear encoder $\bgamma$ in the presence of a complex nuisance $\bX$ presents a formidable challenge. Standard unsupervised methods, such as Principal Component Analysis (PCA), construct encoders by maximizing the marginal variance of $\bZ$ and ignoring $Y$ entirely; they therefore capture dominant technical or biological variation in $\bZ$ but not necessarily the variation that is predictive for $Y$ conditional on $\bX$. On the other hand, classical Sufficient Dimension Reduction (SDR) methods, such as Sliced Inverse Regression (SIR) \citep{li1991sliced}, target predictive directions but typically do not exploit the conditional structure of $Z$ given $X$ and cannot handle multi-modal data. These methods often rely on slicing or smoothing over the covariate distribution, which suffers from the curse of dimensionality and becomes technically intractable when $X$ contains mixed discrete–continuous variables like CNAs \citep{chiaromonte2002sufficient}. Furthermore, modern "black-box" explainability methods applied post hoc to complex models and AI-based encoders often fail to provide faithful explanations of the underlying mechanism \citep{rudin2019stop}.

Therefore, a new framework is required to bridge this gap. We seek an encoder that is (i) {supervised}, directly utilizing $Y$ to identify predictive signals; (ii) {conditional}, built to encode $\bZ$ specifically after adjusting for the nuisance $\bX$; and (iii) {white-box}, where the weights $\bgamma$ arise from explicit estimation rather than opaque training. In our application, such an encoder $\bgamma^\top \bZ$ would function as an $\bX$-adjusted genetic risk score, capturing the incremental signal of gene expression beyond what is already explained by clinical factors. This conditional perspective is central to our study: it ensures that the learned direction captures novel genetic mechanisms rather than re-encoding known clinical effects. In the following section, we will rigorously define this encoder based on the conditional mean structure of the data and introduce a method based on Stein's identity to estimate it.

\section{Methodology: The Stein-Encoder}\label{sec:methodology}


In this section, we begin by defining the encoder in a general population sense via conditional Stein kernel and providing an interpretation in Section 3.1. Then, we provide the specific choice of the Stein-Encoder to strike a balance between interpretability and tractability in Section 3.2. In the end, we proceed to derive tractable closed-form estimators under a conditional linear-Gaussian structure, present a unified algorithm for nuisance removal and index recovery, and show how we use the estimated Stein-Encoder in the METABRIC study.

\subsection{Stein-Encoder: Definition and Interpretation}\label{sec:stein-definition}
Let $\mathcal{H}_k(\bZ\mid \bX)$ be the $k$-th order conditional Stein kernel defined using the conditional density $p(\bz \mid \bx)$ of $\bZ$ given $\bX=\bx$, the specific form can be found in \citep{stein1981estimation}. The $k$-th order Stein-Encoder is defined as a projection of $Z$ over the direction of the rank-one decomposition of the $k$-th Stein moment tensor $\mathcal{M}_k(T) := \mathbb{E}[\,T(Y)\,\mathcal{H}_k(\bZ\mid \bX)\,]$ as stated in Definition \ref{def:pop_stein_encoder_k} where $T$ is a probe function. Such a definition involves $Y$, $\bX$ and $\bZ$, which can be viewed as an association strength between $Y$ and $\bZ$ conditional on $\bX$, and is thus different from PCA and SIR.


\begin{definition}[Stein-Encoder]
\label{def:pop_stein_encoder_k}
The $k$-th order Stein-Encoder is the scalar index:
\[
\mathrm{Enc}_k(\bZ) := \bgamma^\top \bZ.
\]
where $\bgamma$ is the rank-one decomposition of the $k$-th order Stein moment tensor $\mathcal{M}_k(T) := \mathbb{E}[\,T(Y)\,\mathcal{H}_k(\bZ\mid \bX)\,].$
\end{definition}

Under mild conditions, we can show that the Stein-Encoder sufficiently summarizes the impact of $\bZ$ on $Y$ after adjusting for $\bX$ and can be interpreted as a statistically $\bX$-adjusted score of $\bZ$ in predicting $Y$. In specific, consider the case where the relevant information in $\bZ$ for predicting $Y$ (conditional on $\bX$) is captured entirely by a single linear projection $\bbeta^\top \bZ$ for some unknown direction $\bbeta \in \mathbb{R}^q$. Mathematically, this implies the conditional expectation satisfies: $\mathbb{E}[Y \mid \bX, \bZ] = \mathbb{E}[Y \mid \bX, \bbeta^\top \bZ].$
\textcolor{black}{Under this single-index assumption, the general data generating process $Y = g(\bX, \bZ) + \varepsilon$ introduced in \eqref{eq:general_model} simplifies to the multi-modal single-index model summarizing the impact of the modality of interest into a single value:
\begin{equation}
    \label{eq:si_model}
    Y = f(\bX, \bbeta^\top \bZ) + \varepsilon,
\end{equation}
where $f$ is an unknown link function.} Crucially, we claim that under this structural assumption, the Population Stein-Encoder $\bgamma$ defined in Definition \ref{def:pop_stein_encoder_k} perfectly aligns with the true structural direction $\bbeta$ (up to a sign flip). Therefore, the Stein-Encoder bears a statistical interpretation of $\bX$-adjusted score of $\bZ$ in predicting $Y$

To understand why this alignment holds, we examine the components of the Stein moment $\mathcal{M}_k(T)$. The core engine is the conditional Stein kernel $\mathcal{H}_k(\bZ\mid \bX)$. For the first order ($k=1$), this kernel is simply the negative conditional score function, $\mathcal{H}_1(\bZ\mid \bX) = -\nabla_{\bz} \log p(\bZ\mid \bX)$. Intuitively, weighting the features by this score function analytically ``residualizes'' $\bZ$ against $\bX$. Under the model \eqref{eq:si_model} and standard regularity conditions, a generalized integration by parts argument yields the Stein identity:
\begin{equation}
\label{eq:stein_identity}
\mathcal{M}_k(T) = \mathbb{E}\left[ \frac{\partial^k \,\mathbb{E}\{T(Y)\mid \bX,\bZ\}}{\partial (\bbeta^\top \bZ)^k} \right] \cdot \bbeta^{\otimes k}.
\end{equation}
This identity reveals that the observable moment tensor $\mathcal{M}_k(T)$ is proportional to $\bbeta^{\otimes k}$. Therefore, the vector $\bgamma$ recovered from the decomposition must satisfy $\bgamma = \pm \bbeta$, ensuring that the Stein-Encoder correctly identifies the predictive index required for the model.

In the context of our cancer application, the Stein-Encoder $\bbeta^\top \bZ$ serves as an $\bX$-adjusted genetic risk score. Unlike PCA, which captures marginal variation in $\bZ$ regardless of $Y$, or standard neural networks, which entangle $\bX$ and $\bZ$, the Stein-Encoder explicitly isolates the genetic signal predictive of $Y$ that is orthogonal to the clinical factors $\bX$. By leveraging the conditional score function, the method effectively filters out the nuisance variation explained by $\bX$, ensuring that the learned direction reflects valid biological heterogeneity rather than confounding clinical effects. 


The practical implementation of the Stein-Encoder requires navigating a fundamental trade-off between interpretability and estimability across three key design elements: (1) the \textbf{order} $k$, where higher orders ($k \ge 3$) theoretically capture complex dependencies but involve the statistical challenge of estimating high-dimensional tensors; (2) the \textbf{probe function} $T(Y)$, which targets specific functionals of the conditional distribution but requires careful selection to prevent identification failure due to vanishing coefficients; and (3) the \textbf{Stein kernel} $\mathcal{H}_k(\bZ\mid \bX)$, where flexible nonparametric estimators of the conditional density $p(\bz\mid \bx)$ maximize generality but suffer from instability in high dimensions compared to structured working models. Allowing arbitrary $k$, $T$, and $p(\bz\mid \bx)$ makes the interpretation of $\mathrm{Enc}_k(\bZ)$ extremely broad (it becomes a general probe of conditional sensitivity), but estimation becomes infeasible. In practice, we need to restrict them appropriately to yield encoders that are both interpretable and accurately estimable in
realistic sample sizes for the study of interest. The specific choice of the order, probe and stein kernel for the METABRIC study will be specified in Section \ref{Sec:choice}.

\subsection{Stein-Encoder: Choice for METABRIC study}\label{Sec:choice}

In this subsection, to balance interpretability and tractability in the METABRIC analysis, we adopt a structured configuration defined by three specific choices: (1) \textbf{Order:} we restrict attention to $k \in \{1, 2\}$, which is sufficient to capture dominant mean and volatility effects while avoiding the estimation instability associated with higher-order tensors; (2) \textbf{Probe Dictionary:} we utilize a versatile set of transformations $\mathcal{T} = \{y, y^2, \arctan(ay), \frac{ay^2}{1+ay^2}\}$, which is theoretically guaranteed to identify complex dependencies under the conditions established in Section C of the Supplementary Materials; and (3) \textbf{Conditional Stein Kernel:} we impose a conditional linear-Gaussian working model $\bZ\mid \bX \sim \mathcal{N}(\bA \bX, \bSigma)$, which simplifies the score function to a linear transformation of the residuals $\bZ-\bA \bX$, thereby enabling closed-form estimation and explicitly isolating the genetic signal orthogonal to the clinical background.
Motivated by the need to explicitly disentangle the high-dimensional features $\bZ$ from the complex
nuisance covariates $\bX$ in a statistically stable way, we impose the following working model for
$\bZ \mid \bX$.

\begin{assumption}[Conditional Linear-Gaussian]
\label{assump:cond_gaussian}
Conditional on $\bX$, the vector $\bZ$ follows a multivariate Gaussian distribution:
\begin{equation}
\label{eq:cond_gaussian}
\bZ \mid \bX \sim \mathcal{N}(\bA \bX,\bSigma),
\end{equation}
where $\bA \in \mathbb{R}^{q\times p}$ is an unknown coefficient matrix capturing the baseline effect
of $\bX$ on $\bZ$, and $\bSigma\succ 0$ is an unknown covariance matrix.
\end{assumption}

Under Assumption~\ref{assump:cond_gaussian}, the score function becomes explicit: $\mathcal{H}_1(\bZ\mid \bX) = -\bSigma^{-1}(\bZ - \bA \bX)$. Plugging this into the general definition yields the following closed-form residual Stein moment tensor:
\begin{align}
    \label{eq:resid_stein_identities}
    \text{1st Order:}\quad & \mathbb{E}\bigl[T(Y)\bSigma^{-1}(\bZ-\bA \bX)\bigr]  \\
    \text{2nd Order:}\quad & \mathbb{E}\Bigl[T(Y)\bigl\{\bSigma^{-1}(\bZ-\bA \bX)(\bZ-\bA \bX)^\top\bSigma^{-1} - \bSigma^{-1}\bigr\}\Bigr], \nonumber
\end{align}
Under the multi-modal single-index model \eqref{eq:si_model}, these residual Stein moments exhibit a direct structural alignment with the true index $\bbeta$. Specifically, the first-order moment is proportional to $\bbeta$, while the second-order moment is proportional to the rank-one matrix $\bbeta\bbeta^\top$. Consequently, the Stein-Encoder $\bgamma$—obtained by standardizing the first-order vector or computing the leading eigenvector of the second-order matrix—satisfies $\bgamma = \pm \bbeta$. This guarantees that the Stein-Encoder consistently identifies the genetic direction of interest; detailed derivations are provided in Section C of the Supplementary Materials.

\textcolor{black}{Assumption 3.1 is compatible with the METABRIC study, where the raw gene-expression measurements were converted to relative expression \(z\)-scores using the diploid (\(\mathrm{CNA}=0\)) reference sample. After this processing, the marginal distribution of each gene is approximately standard normal by construction, which is further illustrated by Figure 3 in Section~B.5 of the Supplementary Materials. Moreover, the underlying Stein construction in Definition \ref{def:pop_stein_encoder_k} only requires the conditional score \(s(\mathbf{z},\mathbf{x})=\nabla_z \log p(\mathbf{z}\mid \mathbf{x})\), so more general conditional models beyond Gaussian, such as conditional elliptical and generalized linear models, are possible in principle. However, a more general and flexible choice of the working model \(p(\mathbf{z}\mid \mathbf{x})\) is  typically at the cost of a more complex stein identity and a nonlinear score structure, which requires further study. The Gaussian case, besides practically relevant to the METABRIC study, is methodologically attractive with its simple closed-form estimators. Under correct specification, this structure supports the identification result \(\boldsymbol{\gamma}=\pm\boldsymbol{\beta}\) and the generalization theory developed in Section 4. Under misspecification, the theoretical robustness and empirical improvements are still achieved as illustrated in Section B.3.}

Crucially, both identities depend only on low-order moments of the {residuals} $\bZ-\bA \bX$ and
the precision matrix $\bSigma^{-1}$; they do not require estimating the unknown link function $f$.
Under Assumption~\ref{assump:cond_gaussian}, this means that the population Stein-Encoder can
be fully characterized in terms of $\bX$-residualized, whitened gene expression. In our application, the resulting index $\bgamma^\top \bZ$ can thus be interpreted as an $\bX$-adjusted, linearly weighted genetic risk score for the transformed outcome $T(Y)$, recovered from simple residual moments rather than from a black-box predictive model.

\subsection{Stein-Encoder: Estimation and Use}\label{sec:estimation_algo}

In this subsection, we propose an algorithm to estimate the Stein-Encoder tailored for the METABRIC study in Section \ref{Sec:choice}, and show how to use the estimated Stein-Encoder to help achieve the prediction and interpretation tasks considered in the METABRIC study. The estimation of the Stein-Encoder mainly consists of the following two steps, (1) estimate the Stein kernel and (2) select the probe and the order, as summarized in Algorithm \ref{alg:unified_stein_encoder_concise}.

\begin{algorithm}[H]
\caption{Unified Sequential Stein-Encoder}
\label{alg:unified_stein_encoder_concise}
\begin{algorithmic}[1]
\State \textbf{Input:} Data $(\bY, \bX, \bZ)$; Ordered Probe List $\mathcal{T} = \{T_1,... T_4\}$; Threshold $\tau_1$, $\tau_2$.
\State \textbf{Step 1: Nuisance Removal}
\State \quad Estimate $\widehat{\bA}$ and $\widehat{\bOmega}$ via Algorithm 1 in Supplementary materials. Compute residuals $\widehat{\bZ}'_i = \bZ_i - \widehat{\bA} \bX_i$.
\State \textbf{Step 2: Sequential Identification}
\For{each probe $T \in \mathcal{T}$}
    \State \textbf{Attempt First-order ($m=1$):}
    \State \quad Compute whitened Stein vector $\widehat{\bnu} = \widehat{\bOmega} \left( \frac{1}{n} \sum_{i} T(Y_i) \widehat{\bZ}'_i \right)$.
    \State \quad If signal strength $\|\widehat{\bnu}\|_2 > \tau_1$, \textbf{go to Step 3} with $(m=1, \widehat{\bu}=\widehat{\bnu})$.
    \State \textbf{Attempt Second-order ($m=2$):}
    \State \quad Compute Stein matrix $\widehat{\bK} = \widehat{\bOmega}^{1/2} \left[ \frac{1}{n} \sum_{i} T(Y_i) (\widehat{\bZ}'_i \widehat{\bZ}'_i{}^\top - \widehat{\bSigma}) \right] \widehat{\bOmega}^{1/2}$.
    \State \quad Compute leading eigenpair $(\lambda, \bv)$.
    \State \quad If signal strength $|\lambda| > \tau_2$, \textbf{go to Step 3} with $(m=2, \widehat{\bu}=\widehat{\bOmega}^{1/2}\bv)$.
\EndFor
\State \textit{Fallback:} Select $(m^*, \widehat{\bu}^*)$ corresponding to the maximum signal strength observed.
\State \textbf{Step 3: Recovery}
\If{\textit{High-Dim Regime}}
    \State \quad $\widehat{\bgamma}_0 \leftarrow \text{Trunc}(\widehat{\bu})$ (Hard-thresholding for $m=1$; Truncated Sparse PCA \citep{yuan2013truncated} output for $m=2$).
\Else
    \State \quad $\widehat{\bgamma}_0 \leftarrow \widehat{\bu}$.
\EndIf
\State \Return Normalized estimator $\widehat{\bgamma} = \widehat{\bgamma}_0 / \|\widehat{\bgamma}_0\|_2$.
\end{algorithmic}
\end{algorithm}

In Algorithm 1, we first calculate the residual $\widehat{Z}'_i = Z_i - \widehat{A}X_i,\; i=1,\dots,n$, with the estimated conditional mean matrix $\widehat{\bA}$ and precision matrix $\widehat{\bOmega}=\widehat{\bSigma}^{-1}$. Step 1 aims to facilitate the calculation of the Gaussian Stein kernel later. Then, we proceed with a sequential selection of the probe and order. In specific, for each probe $T$ considered in Section \ref{Sec:choice}, we calculate the signal strength by the first order whitened Stein vector, and select the order $m=1$ when the signal is strong enough and $m=2$ otherwise to remedy the potential identifiability issue under first order Stein-Encoder. The selected pair of probe and order is the one achieving the maximum signal strength. In the end, we estimate the Stein-Encoder by hard thresholding when the selected order is 1 or truncated PCA otherwise. We also provide a detailed algorithm in Section A of the Supplementary materials.

Given the estimated Stein-Encoder $\widehat{t}=\widehat{\bgamma}^\top \bZ$, we can improve prediction accuracy in multi-modal studies and interpret the impact of the modality of interest. First, we can use the scalar $\hat{t}$ to interpret the impact of the modality of interest. In specific, rather than black-box encoders, $\hat{t}$ can be interpreted as the conditional genetic risk score orthogonal to the clinical baseline in the METABRIC study.  Second, we can 
feed $(\bX,\widehat{t})$ into downstream DNN to improve prediction of $Y$ in the multi-modal study. In specific, rather than compressing both clinical modality $\bX$ and genetic modality $\bZ$, we first compute the Stein conditional encoded genetic risk scores $\widehat{t}=\widehat{\bgamma}^\top \bZ$ and then
train a multi-layer perceptron $h_\Theta(\bX,\widehat{t})$ to predict cancer-related outcomes $Y$ in the METABRIC study (with an optional residual step in Eq. (3.24) of the Supplementary Materials to safeguard robustness). \textcolor{black}{This ``encoder + downstream'' strategy decouples directional recovery from joint link estimation and substantially reduces the optimization burden, while directly fitting a joint training model might suffer from issues of identification, complex optimization and nuisance error as illustrated and discussed in Section B.1 of the Supplementary material.} In summary, based on the Stein-Encoder, the two-stage prediction can substantially improve prediction accuracy of clinical outcomes because it mitigates the curse of dimensionality by filtering out high-dimensional genomic noise while preserving the interpretability and specific signal relevant to the outcome. 

In the two-stage prediction of the METABRIC study, we do not further compress $\bX$ because it consists of clinical and CNA variables which include many numerical variables and are difficult to
interpret using parametric structure. In addition, we can further augment this two-stage architecture with a residual learning safeguard that
fits an additional network on $(\bX,\bZ)$ to model the remaining signal beyond $(\bX,\widehat{t})$, so
that the combined predictor enjoys a provable safety guarantee when the genetic impact of the METABRIC study might not be linear. The precise formulation and risk bound is in Section C of the Supplementary Material.

\section{Theoretical Properties}\label{sec:theory}

In this section we establish the main theoretical guarantees of the Stein-Encoder. We proceed
in three parts. In Section~\ref{sec:probe_identification} we analyze when simple probe choices
fail to identify the index direction and develop a principled probe dictionary that guarantees
identification under mild conditions. Section~\ref{sec:consistency} then studies the statistical
properties of the estimator: we show that the unified algorithm in
Section~\ref{sec:estimation_algo} yields consistent estimators of $\gamma$ in both low- and
high-dimensional regimes. Finally, Section~\ref{sec:nn_benefits} quantifies how using the
estimated Stein-Encoder as an input to downstream neural networks improves their generalization
performance compared to naive models that operate directly on $(\bX,\bZ)$.

\subsection{Probe Design and Identification}\label{sec:probe_identification}

The estimation framework in Section~\ref{sec:estimation_algo} recovers $\gamma$ from sample Stein moments provided that, at the population level, at least one Stein coefficient is non-zero: $c_1(T;a)\neq 0$ or $c_2(T;a)\neq 0$ for some probe $T(\cdot;a)$ in the dictionary. If we choose probes poorly, for example by using only the identity $T(Y)=Y$ or the square $T(Y)=Y^2$, these coefficients can vanish even when $\gamma\neq 0$ because of the specific structure of the link $f$ in the multi-modal single-index model $Y = f(\bX,\bbeta^\top \bZ) + \varepsilon.$ We explicitly analyze these degeneracy phenomena—specifically symmetry (parity) failure and high-frequency orthogonality—and provide detailed mathematical constructions in Section C of the Supplementary Materials.

These potential failures show that relying only on $T_1(Y)=Y$ and $T_2(Y)=Y^2$ is not sufficient: we need a small dictionary of probes that
(i) handles common linear and variance type relationships efficiently,
(ii) has enough “spectral richness” in $Y$ to interact with higher-order structure in the residual index, and
(iii) remains bounded so that the corresponding Stein moments are statistically stable. 

Guided by these principles, we work with the following dictionary:
\[
    T_1(y)=y,\quad
    T_2(y)=y^2,\quad
    T_3(y;a)=\arctan(ay),\quad
    T_4(y;a)=\frac{a y^2}{1+a y^2},\quad a>0.
\]
The first two probes capture linear and volatility-type effects. The bounded odd probe $T_3$ has a Taylor expansion involving $y^3,y^5,\dots$, and the bounded even probe $T_4$ involves $y^4,y^6,\dots$. By varying the scale parameter $a$, these probes generate a wide range of higher-order moment patterns in $Y$, while keeping all moments finite. In particular, $T_3$ and $T_4$ can interact with higher-order components of $f(\bX,\bbeta^\top Z)$ that are orthogonal to the linear and quadratic structure picked up by $T_1$ and $T_2$.

Formally, under the conditional Gaussian model for $\bZ\mid \bX$ and mild regularity and non-degeneracy assumptions on the conditional moments of $Y$ given the residual index (see Assumptions 2, 3 in Section C of the Supplementary Materials), one can show that this dictionary is rich enough to avoid the degeneracies described above.

\begin{theorem}[Unified identification with probe dictionary]
\label{thm:unified_identification}
Suppose Assumption~\ref{assump:cond_gaussian} holds and the Stein regularity and non-degeneracy conditions in Assumptions 2 and 3 in Section C of the Supplementary Materials are satisfied. Then, under the multi-modal single-index model \eqref{eq:si_model}, there exists at least one probe $T\in\{T_1,T_2,T_3(\cdot;a),T_4(\cdot;a)\}$, an order $j\in\{1,2\}$, and (for $T_3,T_4$) a sufficiently small $a>0$ such that the corresponding Stein coefficient is non-zero: $c_j(T;a)\neq 0$. Equivalently, $\bbeta$ is identified via either the first order or the second order Stein-Encoder $\bgamma$ using a probe from this dictionary ($\bgamma = \pm \bbeta$).
\end{theorem}

Theorem~\ref{thm:unified_identification} formalizes the role of the probe dictionary used by our sequential Stein-Encoder algorithm. It guarantees that, under mild assumptions on the single-index model $Y=f(\bX,\bbeta^\top \bZ)+\varepsilon$, scanning over $T\in\{T_1,T_2,T_3,T_4\}$ and orders $j\in\{1,2\}$ will always yield at least one non-degenerate Stein moment and hence an identifiable encoding direction. In practice, the algorithm in Section~\ref{sec:estimation_algo} implements this strategy by evaluating simple signal-strength criteria to select a probe order pair from the data.

\subsection{Consistency of the Stein-Encoder}\label{sec:consistency}

We establish the estimation consistency of the Stein-Encoder in two regimes. We begin with the \textbf{low-dimensional regime}, where $p, q$ grow more slowly than $n$ ($p/n, q/n \to 0$). We utilize the estimators $(\widehat{\bA},\widehat{\bSigma},\widehat{\bOmega})$ from the low-dimensional version of Algorithm 2 in Section A of the Supplementary Materials (OLS and sample covariance). For convenience, define the aggregate rate $r_n := \sqrt{(p+q)/n}$.

\begin{theorem}[Low-Dimensional Consistency]
\label{thm:lowdim_consistency}
Suppose Assumptions \ref{assump:cond_gaussian}, 2 and 3 in Section C of the Supplementary Materials hold. Let $(k, T)$ be the order and probe used in Algorithm~\ref{alg:unified_stein_encoder_concise} and $\widehat{\bgamma}^{(k)}$ computed by Algorithm \ref{alg:unified_stein_encoder_concise}. Provided the identification condition holds (i.e., $c_k(T;a) \neq 0$), there exists a sign $l\in\{-1,1\}$ such that the estimator $\widehat{\bgamma}^{(k)}$ satisfies:
\[
    \bigl\|\widehat{\bgamma}^{(k)} - l\,\bgamma\bigr\|_2 = O_p(r_n).
\]
\end{theorem}
Theorem \ref{thm:lowdim_consistency} confirms that in standard settings, $\widehat{\bgamma}$ converges at the parametric rate adapted to the joint dimension. Next, we consider the \textbf{high-dimensional regime}, where $p, q$ may exceed $n$. Here, we assume $\bgamma$ is $s_\gamma$-sparse, $\bA$ is $s_A$-sparse, $\bOmega$ is $s_{\Omega}$-sparse and utilize the high-dimensional version of Algorithm~\ref{alg:unified_stein_encoder_concise} (Lasso and Graphical Lasso) with truncation level proportional to $s_\gamma$. The nuisance estimators $(\widehat{\bA}, \widehat{\bOmega})$ are calculated by Algorithm 2 in Section A of the Supplementary Materials, and they can be proved to satisfy the standard high-dimensional convergence rates detailed in Section C of the Supplementary Material.

\begin{theorem}[High-Dimensional Consistency]
\label{thm:hd_consistency}
Suppose Assumptions \ref{assump:cond_gaussian} and 2–4 in Section C of the Supplementary Materials hold. Let $(k, T)$ be the order and probe used with truncation level proportional to $s_\gamma$ in Algorithm \ref{alg:unified_stein_encoder_concise} and $\widehat{\bgamma}^{(k)}$ computed by Algorithm \ref{alg:unified_stein_encoder_concise}. Provided the identification condition holds (i.e., $c_k(T;a) \neq 0$), there exists a sign $l\in\{-1,1\}$ such that the sparse estimator $\widehat{\bgamma}^{(k)}$ satisfies:
\[
    \bigl\|\widehat{\bgamma}^{(k)} - l\,\bgamma\bigr\|_2
    = O_p\Bigl(
        \sqrt{\frac{s_\gamma \log q}{n}}
        + \sqrt{\frac{s_A \log(p\vee q)}{n}}
        + \sqrt{\frac{s_\Omega \log q}{n}}
      \Bigr).
\]
\end{theorem}

Comparing the two regimes, the Stein-Encoder exhibits a unified behavior: whether relying on the first-order mean signal or the second-order curvature signal, the estimation error is consistently governed by the sum of the complexity of estimating the target index (represented by $r_n$ or $\sqrt{s_\gamma \log q/n}+\sqrt{{s_A \log(p\vee q)}/{n}}+ \sqrt{{s_\Omega \log q}/{n}}$) and the cost of estimating the nuisance background. The high-dimensional rate explicitly accounts for the sparsity of the genetic signal ($s_\gamma$) and the complexity of the clinical/background associations ($s_A, s_\Omega$), ensuring the method remains rigorous even when $p, q \gg n$.

\subsection{Benefits for Downstream Neural Networks}\label{sec:nn_benefits}
\textcolor{black}{A natural question is whether and by how much the Stein-Encoder's dimensionality reduction translates into a provable improvement in
downstream prediction. In this section, we answer this question affirmatively by comparing the excess prediction risk of a naive MLP fitted on \((\mathbf{X},\mathbf{Z})\) (Proposition~\ref{Naive-Gen-Bound}) with that of an MLP fitted on the reduced input \((\mathbf{X},\widehat{\bgamma}^\top \mathbf{Z})\) (Theorem~\ref{thm:nn-bound}), and show that the latter delivers a substantial risk reduction, particularly in the METABRIC study.} 

In particular, here we consider the feedforward neural network as the regression model of the downstream prediction task, $ f_{\mathbf\Theta}(\mathbf{X},\mathbf{Z})= \Theta^L \boldsymbol{f}^L\left[\Theta^{L-1} \cdots \boldsymbol{f}^1\left[\Theta^0 (\mathbf{X},\mathbf{Z})\right]\cdots\right]$, indexed by the parameter space $\mathcal{M}=\{\mathbf\Theta = (\Theta^L,\dots,\Theta^0),\Theta^l \in \mathbb{R}^{p_{l+1} \times p_l},l\in[L],p_0 = p+q, p_{L+1} = 1\}$. The functions $f^l : \mathbb{R}^{p_l} \to \mathbb{R}^{p_l}$ are 1-Lipschitz activation functions, $p_0 = p+q$ is the input dimension, $p_{L+1} = 1$ is the output dimension, $L$ is the network depth, $p_{\max} = \max_{l \in \{0,\ldots,L-1\}} p_{l+1}$ is the maximum width, and $p_{\operatorname{total}} = \sum_{l=0}^L p_{l+1} p_l$ is the total number of parameters. Without loss of generality, similar to \citep{lederer2024statistical} we focus our analysis on the constrained parameter spaces $\mathcal{M}_1 = \left\{\mathbf\Theta \in \mathcal{M} : \max_{l \in \{0,\ldots,L-1\}} \|\Theta^l\|_{1,1} \leq 1\right\}$ where $\|\Theta^l\|_{1,1} = \sum_{i=1}^{p_{l+1}} \sum_{k=1}^{p_l} |(\Theta^l)_{ik}|$ and the derived theoretical results can be easily generalized to any parameter space with bounded $l_1$-norm. The naive MLP estimator is defined as $\widehat{\mathbf\Theta}_{0}:=\arg\min_{\mathbf\Theta \in \mathcal{M}_1} \frac{1}{n}\sum_{i=1}^n \left\|Y_i  - f_{\mathbf\Theta}(\mathbf{X}_i,\mathbf{Z}_i)\right\|_2^2 + \lambda_0\|\Theta^L\|_{1,1}$, where the regularization term is proportional to the $l_1$ norm of the last-layer parameters. This is to simplify the theoretical analysis, and the result can be extended to more common and complex regularization terms like $\lambda \sum_{l=0}^L \|\Theta^l\|_{1,1}$ or $\lambda \prod_{l=0}^L \|\Theta^l\|_{1,1}$ using the same analysis technique \citep{lederer2024statistical}. The generalization bound of the neural network with parameter $\widehat{\mathbf{\Theta}}_0$ can be given as follows.
\begin{Proposition}\label{Naive-Gen-Bound}
When $Y - \mathbb E[Y|\mathbf{X},\mathbf{Z}]$ is subgaussian, $\mathbb E[Y|\mathbf{X},\mathbf{Z}]$ is a 1-Lipschitz function of $(\mathbf{X},\mathbf{Z})$ and $\lambda_0\asymp\sqrt{{p_0 L (\log[2np_{\operatorname{total}}])^3}/{n}}$,  with probability at least $1 - 2/n$,
    \begin{align}\label{eq:Naive-Gen-Bound}
       \mathbb{E}|f_{\widehat{\mathbf\Theta}_{0}}(\mathbf{X},\mathbf{Z}) - \mathbb E[Y|\mathbf{X},\mathbf{Z}]|^2 \lesssim &  \underbrace{\inf_{\mathbf{\Theta}\in\mathcal{M}_1} \mathbb{E}|f_{\mathbf{\Theta}}(\mathbf{X},\mathbf{Z}) - 
 \mathbb E[Y|\mathbf{X},\mathbf{Z}]|^2}_{\mathcal{E}_1}\notag\\& +\underbrace{ v_\infty \sqrt{\frac{L(\log[2n p_{\operatorname{total}}])^3}{n}} \|\Theta^{*L}\|_{1,1} +  v_\infty^2 \frac{L(\log[2n p_{\operatorname{total}}])^3}{n}}_{\mathcal{E}_2},
    \end{align}
    where $\mathbf{\Theta}^* = (\Theta^{*L},\dots,\Theta^{*0})=\arg\min_{\mathbf{\Theta}}\mathbb{E}|f_{\mathbf{\Theta}}(\mathbf{X},\mathbf{Z}) - \mathbb E[Y|\mathbf{X},\mathbf{Z}]|^2$ and $v_\infty = \sqrt{\frac{1}{n}\sum_{i=1}^n \|\mathbf{V}_i\|_\infty^2}$, $\mathbf{V}_i = (\mathbf{X}_i^\top,\mathbf{Z}_i^\top)^\top$.
\end{Proposition}

To utilize the knowledge in $\widehat{\bgamma}$ to improve the regression, we propose to set the covariates as $(\mathbf{X},\widehat{\bgamma}^T\mathbf{Z})$ instead of the previous $(\mathbf{X},\mathbf{Z})$. In particular, now the parameter space of the considered neural network $f_{\mathbf{\Theta}}(\mathbf{X},\widehat{\bgamma}^T\mathbf{Z})=\Theta^L \boldsymbol{f}^L\left[\Theta^{L-1} \cdots \boldsymbol{f}^1\left[\Theta^0 (\mathbf{X},\widehat{\bgamma}^T\mathbf{Z})\right]\cdots\right]$ becomes $\mathcal{M}'=\{\mathbf\Theta = (\Theta^L,\dots,\Theta^0),\Theta^l \in \mathbb{R}^{p_{l+1}' \times p_l'},l\in[L],p_0' = p+1, p_{L+1}' = 1\}$. $p_{\max}' = \max_{l \in \{0,\ldots,L-1\}} p_{l+1}'$ is the maximum width, and $p_{\operatorname{total}}' = \sum_{l=0}^L p_{l+1}' p_l'$ is the total number of parameters. For simplicity, we still consider the constrained parameter space $\mathcal{M}_1' = \left\{\mathbf\Theta \in \mathcal{M}' : \max_{l \in \{0,\ldots,L-1\}} \|\Theta^l\|_{1,1} \leq 1\right\}$ and the neural network's parameter with $\widehat{\bgamma}$'s knowledge transferred can be estimated as
\begin{align*}
    \widehat{\mathbf\Theta}_{\widehat{\bgamma}}:=\arg\min_{\mathbf\Theta \in \mathcal{M}_1'} \frac{1}{n}\sum_{i=1}^n \left\|Y_i  - f_{\mathbf\Theta}(\mathbf{X}_i,\widehat{\bgamma}^T\mathbf{Z}_i)\right\|_2^2 + \lambda_1\|\Theta^L\|_{1,1}.
\end{align*}
When the single index model is well specified, the generalization bound of $f_{\widehat{\mathbf\Theta}_{\widehat{\bgamma}}}$ is given in Theorem \ref{thm:nn-bound} below.
\begin{theorem}[Generalization Bound under Correct Specification]
\label{thm:nn-bound}
When $(Y,\mathbf{X},\mathbf{Z})$ satisfies model \eqref{eq:si_model} with the noise $e:=Y- \mathbb{E}[Y|\mathbf{X}, \mathbf{Z}]$ being subgaussian, $\mathbb E[Y|\mathbf{X},\mathbf{Z}]$ is 1-Lipschitz function of $(\mathbf{X},\mathbf{Z})$ and $\lambda_1\asymp\sqrt{{ p'_0 L (\log[2np'_{\operatorname{total}}])^3}/{n}}$, under Assumptions 1 to 3, we have
\begin{align}\label{eq:nn-bound-main}
    &\mathbb{E}\left|f_{\widehat{\mathbf\Theta}_{\widehat{\bgamma}}}(\mathbf{X},\widehat{\bgamma}^T\mathbf{Z}) - \mathbb{E}[Y|\mathbf{X}, \mathbf{Z}]\right|^2 \lesssim \underbrace{\inf_{\mathbf{\Theta}\in\mathcal{M}_1'} \mathbb{E}\left|f_{\mathbf{\Theta}}(\mathbf{X}, \bgamma^T\mathbf{Z}) - \mathbb{E}[Y|\mathbf{X}, \mathbf{Z}]\right|^2}_{\mathcal{E}_1'} \notag\\
    &\quad + \underbrace{ v'_\infty \sqrt{\frac{L(\log[2np'_{\operatorname{total}}])^3}{n}}\|\Theta_{\bgamma}^{*L}\|_{1,1}+  v'^2_\infty \frac{L(\log[2np'_{\operatorname{total}}])^3}{n}}_{\mathcal{E}_2'} + \underbrace{ L \mathbb{E}[\|\mathbf{Z}\|_2^2] \|\widehat{\bgamma} - l\bgamma\|_2^2}_{\mathcal{E}_3'},
\end{align}
with probability at least $1 - 2/n$, where  $\mathbf{\Theta}_{\bgamma}^* := \arg\min_{\mathbf{\Theta} \in \mathcal{M}_1'} \mathbb{E}|f_{\mathbf{\Theta}}(\mathbf{X}, \bgamma^T\mathbf{Z}) - \mathbb{E}[Y|\mathbf{X}, \mathbf{Z}]|^2$ and $v'_\infty = \sqrt{\frac{1}{n}\sum_{i=1}^n \|(\mathbf{X}_i, \bgamma^T\mathbf{Z}_i)\|_\infty^2}$.
\end{theorem}
\textcolor{black}{The bound in Eq.\eqref{eq:nn-bound-main} yields three concrete improvements over Proposition~\ref{Naive-Gen-Bound} when the single-index structure provides a good approximation. First, the Stein-Encoder oracle approximation
error \(\mathcal{E}_1' \le \mathcal{E}_1\), where \(\mathcal{E}_1\) is the oracle approximation error of the neural network class \(\mathcal{M}_1\) for the true function in Proposition~\ref{Naive-Gen-Bound}, since the reduced input class is contained in the full input class. Under \(\beta\)-H\"older
smoothness \citep{yarotsky2017error}, the gap is exponential in dimension: \(\mathcal{E}_1 \lesssim p_{\max}^{-2\beta/(p+q)}\) versus
\(\mathcal{E}_1' \lesssim p_{\max}^{-2\beta/(p+1)}\), where \(p_{\max}\) is
the maximum width defined at the beginning of this section. Second, the Stein-Encoder statistical error \(\mathcal{E}_2' \ll \mathcal{E}_2\), where \(\mathcal{E}_2\) is the finite-sample statistical error of estimating
\(\mathbf{\Theta}^*\) in \(\mathcal{M}_1\) under
Proposition~\ref{Naive-Gen-Bound}, since \(p_{\operatorname{total}}' \ll p_{\operatorname{total}}\)
when \(q\) is large, where \(p_{\operatorname{total}}'\) and
\(p_{\operatorname{total}}\) are the total numbers of parameters of the MLP in our proposed and naive methods. Third, the additional encoding cost \(\mathcal{E}_3' = L\,\mathbb{E}[\|\bZ\|_2^2]\,\|\widehat{\bgamma}-l\bgamma\|_2^2\) introduced by Stein-Encoder is negligible because Theorems~\ref{thm:lowdim_consistency}--\ref{thm:hd_consistency} ensure
\(\|\widehat{\bgamma}-l\bgamma\|_2^2\) converges at a near-parametric rate, faster than \(\mathcal{E}_2'\). In particular, in the METABRIC analysis of Section~6, where \(p=400\), \(q=400\), and
\(n\approx 1{,}900\), the Stein-Encoder reduces the effective input dimension from \(p+q=800\) to \(p+1=401\), which substantially tightens both
\(\mathcal{E}_1'\) and \(\mathcal{E}_2'\), explaining the prediction improvement in Section~6. Similar theoretical results and improvement are provided and discussed when the optional residual safeguard in Eq. (3.24) is imposed in Section~C.8.1 of the Supplementary Materials.}

\textcolor{black}{In this paper, we focus on MLP because the METABRIC data are tabular. However, the underlying dimension-reduction benefit is more general that any downstream learner whose generalization bound deteriorates with input dimension might benefit from replacing $\bZ$ with $\widehat{\bgamma}^\top \bZ$. This is empirically supported by the numerical results in Section B.4 of the Supplementary Materials, where the Stein representation also improves CNN1D and FTTransformer backbones.}

\section{Numerical Experiments}\label{sec:simulation}


We consider a multi-modal single-index setting with a nuisance covariate $\bX \in \mathbb{R}^p$, 
a high-dimensional feature vector $\bZ \in \mathbb{R}^q$, and a scalar response $Y$. 
The covariate $\bX$ is generated from a Gaussian AR(1) model $\bX \sim \mathcal{N}(0, \bSigma_X)$ 
with $(\Sigma_X)_{ij} = \rho_x^{|i-j|}$ and $\rho_x = 0.5$. We examine two configurations for $\bZ$. 
In Feature Setting I (independent), $\bZ$ is generated independently of $\bX$ as 
$\bZ \sim \mathcal{N}(0, \bSigma_Z)$ with $(\Sigma_Z)_{ij} = \rho_z^{|i-j|}$ and $\rho_z = 0.3$. 
In Feature Setting II (correlated), $\bZ$ is conditionally linear in $\bX$: $\bZ = \bA \bX + \bE$. 
For the coefficient matrix $\bA \in \mathbb{R}^{q \times p}$, we consider two scenarios: 
in the low-dimensional setting, each entry of $\bA$ is sampled independently from 
$\mathrm{Unif}(-0.5, 0.5)/\sqrt{p}$. In the high-dimensional setting, $\bA$ is sparse such 
that each row $j$ has exactly $s_A = 10$ non-zero entries (indexed by $S_j$ which is randomly sampled) sampled 
independently from $\mathrm{Unif}(-0.5, 0.5)/\sqrt{s_A}$, while $A_{jk}=0$ for $k \notin S_j$. 
Finally, we set $\bE \sim \mathcal{N}(0, \bSigma_Z)$, so that 
$\bZ \mid \bX \sim \mathcal{N}(\bA \bX, \bSigma_Z)$ and $\bX$ and $\bZ$ are substantially correlated.

The true index direction $\bgamma\in\mathbb{R}^q$ is sparse, with five non-zero entries on
the first few odd coordinates:
\[
    \beta_j=\mathbbm{1} \big( j \in \{ 1,3,7 \}  \big) - \mathbbm{1} \big( j \in \{ 5,9\} \big),
    \qquad
    \bgamma = \bbeta / \|\bbeta\|_2,
\]
and the single index is $t=\bgamma^\top \bZ$, with $\mathbbm{1}(\cdot)$ being the indicator function. Conditional on $(\bX,\bZ)$, the response is
generated as
\[
    Y = f(\bX,t) + \varepsilon,\qquad
    \varepsilon\sim\mathcal{N}(0,\sigma_\varepsilon^2),
\]
where $\sigma_\varepsilon^2$ is chosen so that the signal-to-noise ratio
$\mathrm{SNR}=\operatorname{Var}(f)/\sigma_\varepsilon^2$ equals 5. We consider
three signal specifications:
\[
\begin{aligned}
    \text{Model I:}\quad
    &f(\bX,t)
      = 2\sin(t) + 0.3\,t^2 + 1.3 X_1 - 1.1 X_2 + t X_4,\\
    \text{Model II:}\quad
    &f(\bX,t)
      = t^2 \exp(X_1/2) + 0.5 (X_2^2 - 1) + \sin(X_3),\\
    \text{Model III:}\quad
    &f(\bX,t)
      = \frac{2t}{1 + \exp(-X_1)}
        + \sqrt{|X_2|}\,|t+2|
        + \tfrac{1}{2} t^3
        + \tfrac{1}{2} X_3.
\end{aligned}
\]
Model I is a standard heterogeneous single-index model with polynomial and interaction
terms. Model II is partially symmetric in $t$, so the main signal appears in curvature
rather than the first derivative, a regime where naive first-order Stein methods are known
to fail. Model III mimics a more complex, bio-inspired nonlinear relationship with both
smooth and non-smooth dependence on $(\bX,t)$.

For each model and each feature setting (independent vs.\ correlated), we consider two
dimensional regimes: a low-dimensional regime $(p,q)=(20,20)$ and a high-dimensional
regime $(p,q)=(400,100)$. In every configuration we generate $n_{\mathrm{train}}=2000$
training samples and $n_{\mathrm{test}}=2000$ test samples.

To evaluate the benefit of the Stein-Encoder for downstream learning, we compare three
multi-layer perceptron (MLP) based predictors:
(1) a standard MLP taking the concatenated input $[\bX,\bZ]$ (Method A),
(2) an MLP taking $[\bX,\hat{t}]$ with $\hat{t}=\hat{\bgamma}^\top \bZ$ learned by the
proposed Stein-Encoder with residual safeguard in Eq. (3.24) of the Supplementary Material (Method B),
and (3) an MLP taking $[\bX,\mathrm{PC}(\bZ)]$, where $\mathrm{PC}(\bZ)$ is the first principal
component of $\bZ$ estimated on the training set (Method C). 
All networks share the same architecture: a fully connected feed-forward network with
three hidden layers (128 units per layer), SiLU activation, batch normalization, dropout
regularization, and an output layer for regression. This design ensures that differences in
performance come from the representations of $\bZ$ (raw vs.\ Stein-Encoder vs.\ PCA), not
from differences in network capacity or training.


Table~\ref{tab:model1} summarizes the results over 100 replications. \textcolor{black}{By definition, \(\boldsymbol{\gamma}\) is the normalized index direction and is identifiable only up to sign. We therefore assess recovery using the angle between \(\widehat{\boldsymbol{\gamma}}\) and \(\boldsymbol{\gamma}\), together with the projection loss \(1 - |\langle \widehat{\boldsymbol{\gamma}}, \boldsymbol{\gamma}\rangle|\). Accordingly, the reported ``Angle(Proj)'' gives the angle in degrees, followed by the projection loss in parentheses.}
MSE and $R^2$ are computed on the test set for the three MLP-based predictors
(Method A: $[\bX,\bZ]$, Method B: $[\bX,\hat{\gamma}^\top \bZ]$, Method C: $[\bX,\mathrm{PC}(\bZ)]$).
Across all three models and both dimensional regimes, Method B (Stein-Encoder + MLP)
consistently achieves the lowest test MSE and highest $R^2$. In Model I, where the signal
enters via the mean of $Y$ and the first-order Stein coefficient is non-degenerate,
the Stein-Encoder recovers $\bgamma$ with small angle error and yields clear predictive
gains over the black-box MLP on $[\bX,\bZ]$ (Method A), while the unsupervised PCA encoder
(Method C) performs substantially worse. In Model II, where the dependence on $t$ is
locally symmetric and the signal is primarily of variance/curvature type, the
sequential Stein procedure correctly switches to a second-order Stein moment and still
provides robust recovery of $\bgamma$ and superior prediction, whereas both the naive
MLP and the PCA-based encoder struggle to capture this higher-order structure. In Model
III, which combines complicated nonlinear effects of $(\bX,t)$, the residual Stein-Encoder
continues to isolate the contribution of $\bZ$ conditionally on $\bX$ and remains exceptionally stable under
high dimensionality, while performance of the baselines deteriorates more noticeably as
$(p,q)$ increase. Overall, the simulations confirm the main message of the paper: by
using appropriately chosen Stein probes (first vs.\ second order) and residualization
against $\bX$, the Stein-Encoder produces a low-dimensional, task-relevant summary of $\bZ$
that improves downstream neural prediction without sacrificing interpretability.

\begin{table}[htbp]
\centering
\caption{Simulation results for all models. Each entry is averaged over 100 replications
with $n_{\text{train}} = n_{\text{test}} = 2000$. Angle and projection loss summarize the accuracy of
the estimated index direction $\hat{\bgamma}$. MSE and $R^2$ are computed on the test set.}
\label{tab:model1}
\begin{tabular}{lcccccccccc}
\hline
Model I & $p_X$ & $p_Z$ & Angle(Proj) &
      MSE$_A$ & $R^2_A$ & MSE$_B$ & $R^2_B$ & MSE$_C$ & $R^2_C$ \\
      \hline
      Indep & 20 & 20 &  8.996(0.013) & 1.238 & 0.759 & \textbf{1.101} & \textbf{0.785} & 3.935 & 0.233 \\
      Corr   & 20 & 20 &  7.745(0.010) & 1.294 & 0.757 & \textbf{1.145} & \textbf{0.785} & 4.045 & 0.240 \\
      Indep & 400 & 100 & 11.745(0.021) & 2.987 & 0.413 & \textbf{2.242} & \textbf{0.560} & 5.086 & 0.002 \\
      Corr  & 400 & 100 & 11.570(0.021) & 3.178 & 0.394 & \textbf{2.338} & \textbf{0.554} & 5.167 & 0.015 \\ \hline
Model II \\ \hline
      Indep & 20 & 20 & 11.429(0.021) & 1.653 & 0.693 & \textbf{1.501} & \textbf{0.721} & 4.357 & 0.193 \\
      Corr  & 20 & 20 & 11.049(0.019) & 1.875 & 0.695 & \textbf{1.703} & \textbf{0.724} & 5.102 & 0.171 \\
      Indep & 400 & 100 & 20.662(0.108) & 5.587 & -0.031 & \textbf{3.534} & \textbf{0.354} & 5.647 & -0.042 \\
      Corr  & 400 & 100 & 25.821(0.167) & 5.872 & -0.038 & \textbf{3.844} & \textbf{0.321} & 6.072 & -0.074 \\ \hline
Model III \\ \hline
      Indep & 20 & 20 &  5.022(0.004) & 3.588 & 0.742 & \textbf{3.002} & \textbf{0.784} & 14.746 & -0.059 \\
      Corr  & 20 & 20 &  5.037(0.004) & 4.159 & 0.745 & \textbf{3.515} & \textbf{0.784} & 15.906 & 0.022 \\
      Indep & 400 & 100 & 6.718(0.007) & 6.337 & 0.543 & \textbf{5.057} & \textbf{0.636} & 16.475 & -0.189 \\
      Corr  & 400 & 100 & 6.913(0.007) & 7.323 & 0.538 & \textbf{5.729} & \textbf{0.639} & 18.186 & -0.148 \\
\hline
\end{tabular}
\end{table}

\section{Real Data Analysis: The METABRIC Cohort}\label{sec:realdata}

\subsection{Data Background and Preprocessing}

We apply our Stein-Encoder framework to the METABRIC (Molecular Taxonomy of Breast Cancer International Consortium) cohort data \citep{curtis2012genomic}, a comprehensive, multi-modal, large-scale, clinically annotated collection of primary breast cancer specimens from the UK and Canada. In METABRIC, gene expression profiles are known to be strongly associated with key clinical outcomes, making the proposed supervised representation learning particularly well-motivated. In this work, we leverage METABRIC’s high-dimensional gene expression data (Illumina HT-12 v3 platform) alongside standard clinical covariates to demonstrate how Stein-Encoder can enhance predictive modeling in breast cancer. Specifically, we aim to show that by learning a low-dimensional, task-specific embedding of high-dimensional transcriptomic features ($\bZ$) using the Stein-Encoder improves prediction accuracy when combined with standard clinical covariates ($\bX$).


\textbf{Response ($Y$):} We evaluate robustness across four diverse regression tasks:
\begin{itemize}
    \item \textit{Tumor Size}, a direct morphological measure of disease burden.
    \item \textit{Nottingham Prognostic Index (NPI)}, a composite clinical score combining tumor size, grade, and lymph node status, widely used to stratify prognosis.
    \item \textit{Lymph Node Status}, an indicator of metastatic spread (positive/negative).
    \item \textit{Age at Diagnosis}, which is included as a biological proxy and often correlates with subtype distribution and treatment response patterns in breast cancer.
\end{itemize}
Samples with missing responses were excluded, resulting in approximately $n=1900$ observations. We preprocess and standardize the nuisance covariates $\bX$ and the feature vector $\bZ$.

\begin{itemize}
    \item \textbf{Nuisance $\bX$ (Clinical/CNA):} Includes treatment status, tumor grade, ER status, and Copy Number Aberrations (CNA). Categorical variables were integer-encoded. Variables with missing rates $>30\%$ were dropped.
    \item \textcolor{black}{\textbf{Predictors $\bZ$ (Gene Expression):} High-dimensional gene-expression data from the Illumina HT-12 v3 platform. Following standard preprocessing protocols \citep{lupat2023moanna}, we analyze \(z\)-score transformed relative expression values normalized against diploid samples. We then prescreen the top 400 genes with the highest marginal variance for downstream analysis.}
\end{itemize}


We adopt a 5-fold cross-validation scheme to evaluate the benefits of the Stein-Encoder for initializing downstream neural networks. \textcolor{black}{We also use the same MLP backbone in the METABRIC analysis for consistency with the theory in Section 4.3.} We compare the following architectures, all utilizing the same underlying 3-layer MLP backbone:

\textbf{Standard MLP:} The end-to-end baseline receiving raw features $[\bX, \bZ]$.

\textbf{PCA + MLP:} An unsupervised reduction baseline receiving $[\bX, \text{PCA}_k(\bZ)]$.

\textbf{Stein-Encoder + safeguarded MLP:} We estimate $\widehat{\bgamma}_{Stein}$ using Algorithm 2 on the training fold and feed $[\bX, \widehat{\bgamma}_{Stein}^T \bZ]$ to the network with residual safeguard in Eq. (3.24) in Supplementary Materials.

To comprehensively evaluate predictive accuracy, particularly given the diverse scales and biological interpretations of the chosen response variables, we report the mean squared error (MSE), mean absolute error (MAE) and the coefficient of determination ($R^2$) for each configuration.

\subsection{Results}
\label{subsec:metabric_results}

Table~\ref{tab:metabric_simple} summarizes the 5-fold cross-validation performance of the three competing approaches across four clinical prediction tasks in the METABRIC cohort. Overall, the proposed Stein-Encoder consistently achieves lower MSE and higher $R^2$ compared to both the Standard MLP and the PCA baseline, indicating that a supervised, task-aware compression of transcriptomic features yields more informative low-dimensional representations for downstream prediction.

\begin{table}[htbp]
    \centering
    \caption{5-Fold Cross-Validation Results on METABRIC. The proposed Stein-Encoder consistently achieves lower MSE and higher $R^2$ compared to unsupervised and black-box baselines.}
    \label{tab:metabric_simple}
    \vspace{0.2cm}
    \begin{tabular}{l l c c c}
        \hline
        \textbf{Task} & \textbf{Method} & \textbf{MSE} & \textbf{MAE} & \textbf{$R^2$} \\
        \hline
        \textbf{Tumor Size} & \quad A & 139.924 & 8.613 & 0.075 \\
                            & \quad B & 126.585 & \textbf{7.964} & 0.163 \\
                            & \quad \textbf{C} & \textbf{123.819} & 8.036 & \textbf{0.182} \\
        \hline
        \textbf{NPI Score}  & \quad A & 0.750 & 0.648 & 0.427 \\
                            & \quad B & 0.621 & 0.569 & 0.515 \\
                            & \quad \textbf{C} & \textbf{0.617} & \textbf{0.567} & \textbf{0.529} \\
        \hline
        \textbf{Lymph Nodes}& \quad A & 14.558 & 2.082 & 0.125 \\
                            & \quad B & 13.850 & \textbf{2.049} & 0.168 \\
                            & \quad \textbf{C} & \textbf{13.203} & 2.084 & \textbf{0.206} \\
        \hline
        \textbf{Age}        & \quad A & 66.412 & 6.489 & 0.606 \\
                            & \quad B & 69.058 & 6.636 & 0.590 \\
                            & \quad \textbf{C} & \textbf{63.975} & \textbf{6.298} & \textbf{0.620} \\
        \hline
    \end{tabular}
\end{table}

For the challenging \textit{Tumor Size} prediction task, which is known to exhibit substantial biological and measurement noise, the Standard MLP (Method A) trained directly on $[\bX,\bZ]$ achieves only modest explanatory power ($R^2 \approx 0.075$). The PCA baseline (Method B), which projects the high-dimensional $\bZ$ onto a low-dimensional unsupervised subspace before feeding it into the network, improves performance to $R^2 \approx 0.163$. In contrast, the Stein-Encoder (Method C) further reduces MSE from 139.924 (Standard MLP) and 126.585 (PCA) to 123.819, and increases $R^2$ to approximately 0.182. This suggests that the supervised index $\widehat{\bgamma}_{Stein}^\top \bZ$ captures a more task-relevant genetic component of tumor size than the variance-maximizing directions identified by PCA.

A similar pattern is observed for the \textit{NPI Score}, an integrated prognostic index. The Standard MLP baseline achieves an $R^2$ of roughly 0.427, while PCA improves this to 0.525. The Stein-Encoder attains the best overall performance with the lowest MSE (0.617) and the highest $R^2$ (0.529), reflecting that supervised shrinkage towards the Stein direction can extract a more informative one-dimensional risk score than unsupervised projections, even when signal-to-noise ratio is moderate.

For the \textit{Lymph Nodes Positive} outcome, all three methods achieve relatively modest yet non-trivial predictive performance. Here, PCA already yields a noticeable gain over the black-box baseline ($R^2$ from 0.125 to 0.168), indicating that unsupervised dimensionality reduction mitigates overfitting in this high-dimensional setting. Nevertheless, the Stein-Encoder continues to provide incremental improvements (MSE 13.203 vs.\ 13.850 for PCA, $R^2$ 0.206 vs.\ 0.168), underscoring the benefit of tailoring the representation to the specific regression task.

Finally, for \textit{Age at Diagnosis}, which is a smoother and more strongly structured outcome, all methods achieve higher $R^2$ values compared to the other tasks. The Standard MLP reaches $R^2 \approx 0.606$, PCA marginally reduces this to 0.590 (although slightly increasing MSE), while the Stein-Encoder substantially lowers MSE to 63.975 and increases $R^2$ to approximately 0.620. This highlights that even for relatively ``easy'' tasks with higher signal-to-noise ratios, incorporating a supervised, low-dimensional genetic score can still significantly enhance predictive accuracy.

\begin{figure}[t]
    \centering
    \begin{subfigure}[t]{0.48\textwidth}
        \centering
        \includegraphics[width=\linewidth]{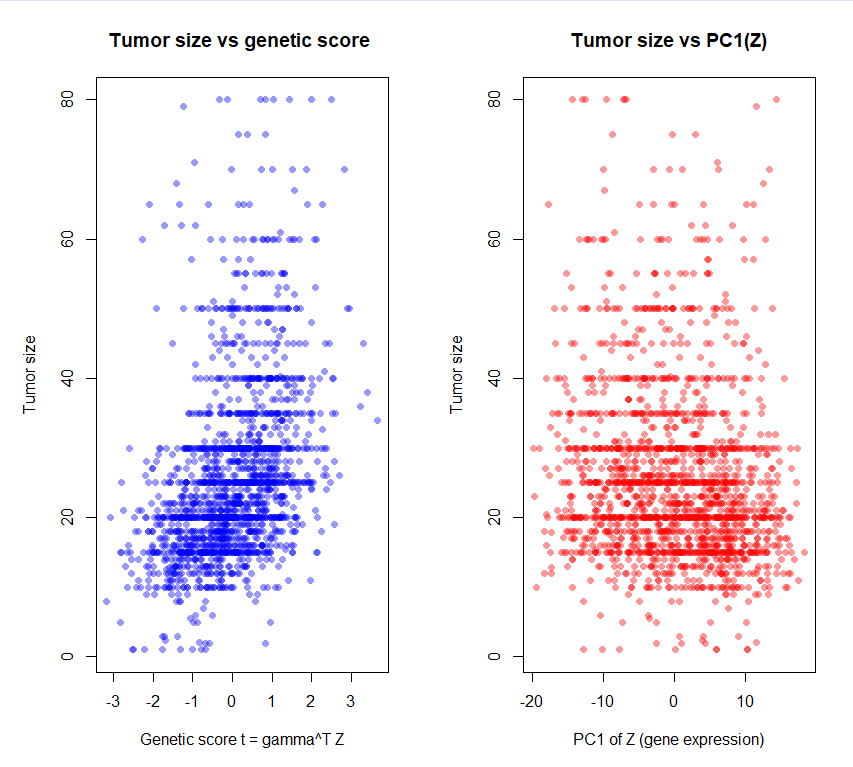}
        \caption{Tumor Size: response vs.\ Stein-Encoder score ($\widehat{\bgamma}^\top \bZ$) and vs.\ first PCA component of $\bZ$.}
        \label{fig:metabric_tumor_scatter}
    \end{subfigure}
    \hfill
    \begin{subfigure}[t]{0.48\textwidth}
        \centering
        \includegraphics[width=\linewidth]{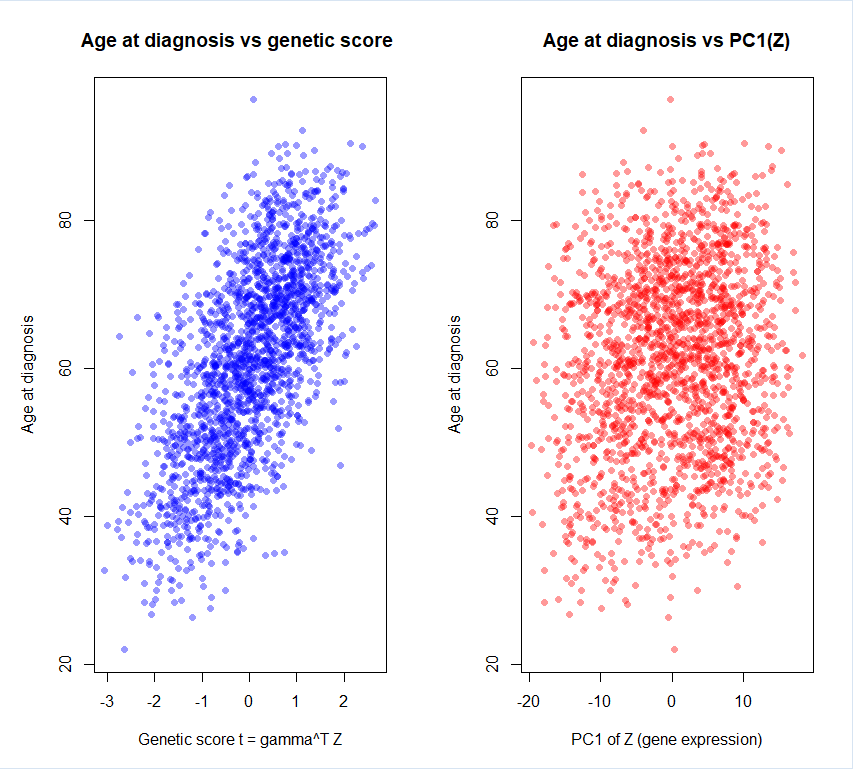}
        \caption{Age at Diagnosis: response vs.\ Stein-Encoder score and vs.\ first PCA component.}
        \label{fig:metabric_age_scatter}
    \end{subfigure}
    \caption{Scatter plots of METABRIC responses versus (i) the supervised Stein-Encoder genetic score $t = \widehat{\bgamma}^\top \bZ$ and (ii) the first principal component of $\bZ$.} 
    \label{fig:metabric_tumor_age_scatter}
\end{figure}

To further visualize the learned genetic representations, we examine the relationship between the response $Y$ and two scalar summaries of the high-dimensional gene expression: the supervised Stein-Encoder score $t = \widehat{\bgamma}_{Stein}^\top \bZ$ and the first principal component $\text{PC1}(\bZ)$ from PCA, where $ \widehat{\bgamma}_{Stein}=\widehat{\bgamma}$ was obtained using the hard-thresholding operator with a sparsity level set $s = 20$. For illustration, we focus on the two tasks where the Stein-Encoder achieves the largest gains over PCA: \textit{Tumor Size} and \textit{Age at Diagnosis}. Figure~\ref{fig:metabric_tumor_age_scatter} displays the corresponding scatter plots for these two responses. \textcolor{black}{Figures for the other two responses, additional normality and single-index structural diagnostics by histogram, QQ plot and residual plot, and the results under alternative downstream architectures (CNN1D and FTTransformer) are reported in Section B.5 of the Supplementary Materials.}

For \textit{Tumor Size} (left panel of Figure~\ref{fig:metabric_tumor_age_scatter}), plotting $Y$ against the Stein-Encoder score reveals a clear, structured monotone trend, indicating that a one-dimensional supervised genetic index explains a substantial portion of the variability in tumor size. In contrast, the scatter of tumor size against the first PCA component appears much more diffuse, with no obvious functional relationship. This illustrates that directions of maximum variance in the gene expression space (PCA) do not necessarily align with directions that are most predictive for the clinical endpoint of interest.

A similar phenomenon is observed for \textit{Age at Diagnosis} (right panel of Figure~\ref{fig:metabric_tumor_age_scatter}). The plot of age versus the Stein-Encoder score exhibits a pronounced and coherent dependence pattern, suggesting that the supervised genetic index successfully captures biologically meaningful age-related transcriptomic variation. By comparison, the relationship between age and the first principal component of $\bZ$ is much weaker and more scattered, again highlighting the limitations of purely unsupervised dimensionality reduction for predictive modeling. Taken together, these visualizations provide direct geometric evidence that the Stein-Encoder is extracting task-specific directions in the gene expression space that are not aligned with principal axes of variance but are substantially more informative for prediction. 


Here, we illustrate that the Stein-Encoder not only improves predictive accuracy but also yields a markedly more interpretable genetic index than PCA. For each of the four prognostic responses (tumor size, lymph nodes examined positive, NPI, and age at diagnosis), the top Stein-Encoder coefficients concentrate on genes that form well-characterized biological modules in breast cancer (Table \ref{tab:genes}), whereas the PCA loadings are dominated by a single, unsupervised proliferation axis that is essentially the same across tasks.

For tumor size and lymph node involvement, the Stein-Encoder upweights core mitotic, cell-cycle, and structure-related genes such as MELK, COL18A1, TNFRSF17, CD1C, TROAP, FOXM1, DLGAP5, CKAP2L, ZWINT, PLK1, CENPA, and CDC20. Many of these (e.g., CDC20, FOXM1, CENPA, CKAP2L, PLK1) belong to or are co-regulated with the mitotic network associated with chromosome 5q deletions and basal-like tumors described by \citep{curtis2012genomic}, which is linked to genomic instability and aggressive disease. That our Stein directions recover highly relevant proliferation and structural programs when predicting tumor size and lymph node involvement is exactly what one would expect if they are capturing residual proliferative aggressiveness beyond standard clinical and CNA covariates.

In contrast, for NPI and age at diagnosis, the Stein-Encoder combines proliferation genes (e.g., CENPE, CDCA5, CCNB2, PTTG1, MCM2) with a rich set of immune and B/T-cell–associated genes, including HLA\mbox{-}DOB, TNFRSF13B, CD52, CD1C, CD48, CD69, IGJ, CD38, AMICA1, RACGAP1, and FOXP1\mbox{-}IT1 (a long non-coding RNA locus related to the FOXP1 tumor suppressor, as highlighted by \citep{pereira2016somatic}). These are canonical markers of B-cell receptor signaling and cytotoxic lymphocytes. \citep{curtis2012genomic} identified a CNA-devoid subgroup with strong adaptive immune signatures and lymphocytic infiltration, and \citep{pereira2016somatic} highlighted FOXP1 as a candidate tumor suppressor whose low expression has been linked to poor outcomes. The fact that our Stein directions for NPI and age are enriched for similar B/T-cell modules is therefore consistent with the ``proliferation vs.\ immune'' prognostic axes reported in METABRIC.

Note that in our implementation, the Stein-Encoder operates on gene expression residuals after adjusting for clinical factors and CNAs, so the selected genes reflect transcriptomic variation not already explained by those covariates. By comparison, the genes with the largest absolute loadings on the first PCA component (e.g., CDCA5, AURKB, UBE2C, CCNB2, CDC20, TPX2, MELK, CEP55, KIF20A, HJURP, PTTG1, CKAP2L, EXO1, BUB1, CDC45, CENPE, NCAPG, CENPA, AURKA, FOXM1) form a generic mitotic module very similar to the one described by \citep{curtis2012genomic}, and this same proliferation component is used for all four responses. This illustrates the key difference: PCA, being unsupervised, always recovers the dominant variance direction (proliferation), whereas the Stein-Encoder produces response-specific directions that align with known, outcome-relevant pathways (proliferation and immune infiltration) and are thus more informative for biological interpretation. 

{\color{black}
In summary, adopting the Stein-Encoder in the METABRIC study, we improve prediction accuracy while producing response-specific genetic scores with clear biological meaning. The improved prediction accuracy supports more accurate diagnosis and timely personalized treatment for breast cancer patients. The interpretable response-specific score helps better understand the mechanism between genes and breast cancers, and identifies significant genes for different breast cancer-related outcomes with clear biological interpretation. 
}

\begin{table}[ht]
\centering
\caption{Top 20 genes selected by the Stein-Encoder for each prognostic response and top 20 genes by PCA}
\label{tab:genes}
\begin{tabular}{ll}
\hline\hline
Response & Stein-Encoder selected genes  \\
\hline
Tumor size &
AMICA1, CD52, MELK, COL18A1, ANKRD50, TNFRSF17, PRF1, CD1C, KIAA0889,\\
& KIF4A, TROAP, DNAJC13, CD79B, FOXM1, DLGAP5, CKAP2L, PMEPA1, ADAM7,\\
& IFNG, C20orf11 \\
\hline
Lymph nodes &
TROAP, ZWINT, SH2D1A, CAMK1G, ZBED2, DDX23, CKAP2L, PLK1, ORAOV1,\\
examined & DPM3, RASGRP2, KIF23, SENP5, CENPA, ARRDC5, TLR10, TNFRSF17, COL18A1,\\
positive & CDC20, HLA\mbox{-}DOB \\
\hline
Nottingham &
ZBED2, CENPE, BANK1, KIFC1, CDCA5, HLA\mbox{-}DOB, TNFRSF13B, POLQ, RASGRP2,\\
Prognostic & PTGER3, CD52, CD1C, CKAP2L, CD48, PLK1, JAK2, JAG2, SIK1, SUSD3, BARD1\\
Index (NPI) &  \\
\hline
Age at &
CD48, MAP4K1, PYHIN1, CCNB2, PTTG1, PRC1, CD69, IGJ, MCM2, IFNG, CD38,\\
diagnosis & ASPM, MELK, RACGAP1, PNOC, FOXP1\mbox{-}IT1, KIF23, AMICA1, ATAD2, NCAPG \\
\hline\hline 
&PCA selected genes (PC1 loadings with the largest absolute coefficients) \\ \hline
& CDCA5, AURKB, UBE2C, CCNB2, CDC20, TPX2, MELK, CEP55, KIF20A, HJURP,\\
& PTTG1, CKAP2L, EXO1, BUB1, CDC45, CENPE, NCAPG, CENPA, AURKA, FOXM1 \\ \hline\hline
\end{tabular}
\end{table}

\section{Significance Statement}\label{sec:discussion}
{\color{black}
The Stein-Encoder is a white-box and supervised method to extract the impact of the modality of interest, while adjusting for the others. Compared with the usual unsupervised or black-box method, the Stein-Encoder makes the prediction more accurate and the interpretation more transparent. In particular, applying the Stein-Encoder pipeline to the METABRIC study, we improve prediction accuracy of the cancer-related outcome and reveal outcome-specific biological patterns, including immune-related signals, supporting more interpretable precision medicine for breast cancer patients.
}

\section*{Acknowledgements} Jiasheng Shi and Xinzhou Guo are co-corresponding authors.

\bibliographystyle{plainnat}
\bibliography{refs}

\end{document}